\documentclass[aps,twocolumn,showpacs]{revtex4}
\usepackage{graphicx}
\usepackage{epsfig,color}
\usepackage{bm}
\begin{document}
\setcounter{page}{1}
\newcommand{\re}[1]{(\ref{#1})}
\newcommand{\lab}[1]{\label{#1}}
\newcommand{\ci}[1]{\cite{#1}}
\renewcommand{\baselinestretch}{1.25}
\newcommand{\bfr}{\begin{flushright}}
\newcommand{\bfl}{\begin{flushleft}}
\newcommand{\efl}{\end{flushleft}}
\newcommand{\efr}{\end{flushright}}
\newcommand{\bc}{\begin{center}}
\newcommand{\ec}{\end{center}}
\newcommand{\be}{\begin{equation}}
\newcommand{\ee}{\end{equation}}
\newcommand{\bea}{\begin{eqnarray}}
\newcommand{\eea}{\end{eqnarray}}
\newcommand{\ba}{\begin{array}}
\newcommand{\ea}{\end{array}}
\newcommand{\nn}{\nonumber}
\newcommand{\edc}{\end{document}}
\newcommand{\ul}{\underline}
\newcommand{\ri}{\rightarrow\infty}
\newcommand{\li}{\leftarrow\infty}
\newcommand{\ra}{\rightarrow}
\newcommand{\la}{\leftarrow}
\newcommand{\ds}{\displaystyle}
\newcommand{\dsf}{\displaystyle\frac}
\newcommand{\dt}{\Delta{t}}
\newcommand{\il}{\int\limits}
\newcommand{\pal}{\partial}
\newcommand{\xxx}{{\it{X}}}
\newcommand{\bone}{{\bf 1}}
\newcommand{\gComment}[1]{}
\renewcommand{\gComment}[1]{\textcolor{red}{Gerardo: #1}}
\title[]{Bose-Einstein Condensate in Weak $3d$ Isotropic Speckle Disorder}

\author{B. \surname{Abdullaev$^{1,2}$}}
\author{A. \surname{Pelster$^{3,4}$}}

\affiliation{
$^1$Institute of Applied Physics, National University of Uzbekistan, Tashkent 100174, Uzbekistan\\
$^2$Institut f\"ur Theoretische Physik, Freie Universit\"at Berlin, Arnimallee 14, 14195 Berlin, Germany\\
$^3$Fachbereich Physik und Forschungszentrum OPTIMAS, Technische Universit\"at Kaiserslautern, 67633 Kaiserslautern, Germany\\
$^4$Hanse-Wissenschaftskolleg, Lehmkuhlenbusch 4, 27753 Delmenhorst, Germany}

\date{Received \today }

\begin{abstract}
The effect of a weak three-dimensional ($3d$) isotropic laser
speckle disorder on various thermodynamic properties of a dilute
Bose gas is considered at zero temperature. First, we summarize the
derivation of the autocorrelation function of laser speckles in $1d$
and $2d$ following the seminal work of Goodman. The goal of this
discussion is to show that a Gaussian approximation of this
function, proposed in some recent papers, is inconsistent with the
general background of laser speckle theory. In this context we also point out that the concept of a quasi-three
dimensional speckle, which appears due to an extension of
the autocorrelation function in the longitudinal direction of a 
transverse $2d$ speckle, is not applicable for the true $3d$
speckle, since it requires an additional space dimension.
Then we propose a
possible experimental realization for an isotropic $3d$ laser
speckle potential and derive its corresponding autocorrelation
function. Using a Fourier transform of that function, we calculate
both condensate depletion and sound velocity of a Bose-Einstein
condensate as disorder ensemble averages of such a weak laser
speckle potential within a perturbative solution of the
Gross-Pitaevskii equation. By doing so, we reproduce the expression
of the normalfluid density obtained earlier within the treatment of
Landau. This  physically transparent derivation shows that
condensate particles, which are scattered by disorder, form a gas of
quasiparticles which is responsible for the normalfluid component.
\end{abstract}

\pacs{ 67.85.Hj,\, 46.65.+g}

\maketitle

\newpage

\section{Introduction}
\label{sec1}

The study of interacting bosonic atoms in a disordered potential
landscape, called in the literature as ``dirty boson
problem``~\ci{Fisher}, has originally been introduced in the context
of the motion of superfluid helium in porous Vycor glass~\ci{Chan}.
Due to the frozen environment, disorder ensembles averages of
physical observables have to be determined, which depend on many
system parameters as, for instance, the strength of a repulsive
interaction between two particles of the Bose gas as well as the
strength and the correlation length which characterize the disorder
potential. The main and intriguing part of the problem is the
competition between the repulsive two-particle interaction and the
localization property of disorder. From a theoretical point of view,
the disorder potential was introduced by investigating the Anderson
localization phenomenon for fermions~\ci{Anderson}. Much attention
has recently been paid for the Anderson localization and the
propagation of bosonic matter waves in random external
potentials~\ci{LSP}. Experimentally, the bosonic matter waves have
been studied in the random potential produced either by laser
speckles~\ci{Billy} or by an incommensurable optical
lattice~\ci{Roati}. Whereas the laser speckle disorder potential is
created by a laser beam scattered from a diffusive glass
plate~\ci{Goodman1}, the incommensurable optical lattice is produced
through two interfering laser beams  with incommensurable
wavelengths. However, one needs to remark that
such lattices exhibit certain
pathological features, which do not occur in genuinely random
lattices, such as a transition between localized and delocalized
states, even in one spatial dimension~\ci{Boers}. In that sense the
quasi-periodic lattices should be considered as to be quasi-random
ones. Recent progress in different experimental realizations of
laser speckle disorder is reported in Refs.~\ci{Kondov,Pezze}.

According to the laser speckle theory described in the seminal
work of Goodman~\ci{Goodman1,Goodman2}, the monochromatic light
reflected from a rough surface on the scale of an optical wavelength
yields many independent dephased but coherent wavelets
which interfere at a distance, which is essentially larger than the
wavelength. This results in a granular pattern of intensity that is
called Gaussian speckle as the real and imaginary parts of the field
amplitude form a circular complex Gaussian distribution at any fixed
spatial point. Details of the speckle formation will be considered
in the next section of the paper. Here, we note that this
distribution consists of the first-order statistics of the speckle
disorder, while the second-order statistics of disorder is
represented by its autocorrelation function.

In order
to understand the underlying physics of laser speckles, let us
briefly describe their formation in $2d$. Object waves are fields,
which are a result of the incident polarized monochromatic field
reflection from a rough surface, and they are described in a plane
$\alpha,\beta$ immediately adjacent to the surface in terms of a complex function
$a(\alpha,\beta)$~\ci{Dainty}. The Huygens-Fresnel principle
establishes in the Fresnel approximation a relation between these object
waves $a(\alpha,\beta)$ and the complex waves $A(x,y)$ in the
observation plane $x,y$ through an integral which resembles a
Fourier transformation. Hence, the wave $A(x,y)$ is a result of the
interference of all object waves in the
$x,y$ plane. As in the Fresnel approximation one assumes the
condition $z \gg (\alpha^2+\beta^2)_{\rm max}/\lambda$, where $z$
denotes the distance between the object wave $\alpha,\beta$ plane as well as
the observation wave $x,y$ plane and $\lambda$ denotes the light wavelength,
the waves $A(x,y)$ are called to be in far field~\ci{Dainty}.

In the Fourier mapping of object waves for the formation of
far fields both the form and the finite size of
the diffraction aperture ${\cal A}$ in the $\alpha,\beta$ plane plays a central role. It
determines the form of the
autocorrelation function as well as its correlation length,
which characterizes the average size of the speckle, i.e.~a grain
of the above mentioned intensity pattern. Typically, the expression
for the autocorrelation function consists of a constant and a
spatially varying part. The latter, which is of interest for various speckle
applications, has one central maximum and
a set of side maxima of decaying height, which are separated from each
other by zeros. This analytical structure is principal in the
theory of laser speckles, since it is the result of the Fourier
transformation of the finite-size diffraction aperture ${\cal A}$.
Due to the existence of zeros, it can qualitatively not be approximated
by a function of a Gaussian form as was assumed and even
numerically derived in Refs.~\ci{Pilati1,Pilati2,Piraud}. It is
interesting that the experiment demonstrates an ambiguity in the following
respect: whereas the function with zeros is exploited in the
papers~\ci{Billy,Clement1a,Clement1b}, the spatial autocorrelation function is fitted
by a Gaussian in Refs.~\ci{Kondov,Pezze,Chen1,Chen2,Chen3,Chen4}. Calculating a standard
deviation of the second-order moment of the random intensity, it was
shown in Ref.~\ci{Clement2} that for $1d$ the autocorrelation
function derived in Ref.~\ci{Goodman1,Goodman2} can be well approximated
by a Gaussian form. However, a Fourier transform of this
autocorrelation function, the power spectral density, which is
essential for the theory of a Bose-Einstein condensate (BEC) in an
external disorder potential, behaves, unlike the Gaussian function,
as the triangle function ${\rm tri}(x)=1-|x|$ for $|x|\leq 1$ and
otherwise zero for any dimensionality. For $1d$ and $2d$ this was
shown by Goodman in Refs.~\ci{Goodman1,Goodman2}, the corresponding $3d$ case is dealt with
below in the text. This triangle function makes the upper limit of
the integration in momentum space finite. For those reasons
the recently proposed Gaussian autocorrelation function for the laser
speckle is not suitable for a comprehensive description  of a BEC
in laser speckle disorder.

The present paper is organized as follows. We start with describing
the basic principles of the laser speckle theory in Sec.~\ref{sec2}.
Following a scheme described in Refs.~\ci{Goodman1,Goodman2}, we
will then derive  in Sec.~\ref{sec3} the expressions for the
autocorrelation function of laser speckles and their Fourier
transforms ranging from $1d$ to $3d$ with special emphasize on
discussing both isotropic and anisotropic cases. The scheme of the
possible experimental realization of the $3d$ isotropic speckle will
be outlined in Sec.~\ref{sec4}.
Note, however, that we consider in our paper a true $3d$ speckle pattern,
not a quasi-three dimensional one of a transverse $2d$ speckle with
a longitudinal depth in the autocorrelation function as described in
Ref.~\ci{Leushacke} and section 4.4.3 of the Goodman book~\ci{Goodman2},
which has been applied in many experiments (see, for instance,
Ref.~\ci{Clement2}). This depth autocorrelation function concept
assumes the existence of an additional spatial direction for the
relevant speckle and can only be valid for $1d$ or $2d$ speckles. As
is further discussed in Refs.~\ci{Leushacke,Goodman2}, the depth size is essentially
larger than ones in other dimensions. Here we consider a $3d$ volume
speckle with compatible speckle grain sizes in all spatial
directions, which was already simulated in
Refs.~\ci{Pilati1,Pilati2}. Since the existing speckle patterns are
experimentally produced mainly in a $2d$ geometry, we will propose a
special scheme for its possible realization in a $3d$ volume. In the
subsequent Sec.~\ref{sec5} the effect of a weak $3d$ isotropic
speckle on various thermodynamic properties of a dilute Bose gas
will be considered at zero temperature. To this end we calculate
both condensate depletion and sound velocity of a BEC within a
perturbative solution of the Gross-Pitaevskii equation. Afterwards,
in Sec.~\ref{sec6}, we reproduce the expression of the normalfluid
density of a BEC in an external disorder potential obtained earlier
within the treatment of Landau. From this rederivation we realize
that condensate particles, which are scattered by a disorder
potential, form a gas of quasiparticles, which is responsible for
the normalfluid component. Finally, we summarize and analyze the
results obtained in the paper in Sec.~\ref{sec7}.

\section{Fundamentals of laser speckle theory}
\label{sec2}

According to Refs.~\ci{Goodman1,Goodman2, Dainty} the circular
Gaussian probability density function
\be p(A_R,A_I)=\dsf{1}{2\pi \eta^2}\exp\left(-\dsf{A_R^2+A_I^2}{2\eta^2} \right)\,,
\lab{lspeck1}
\ee
for the real $A_R$ and imaginary $A_I$ parts of a far-field $A(x,y)$
at each point $x,y$ with the variance $\eta=\sqrt{\langle
|A|^2\rangle}$ represents the background of the theory of laser
speckles. Another basis of the theory is the $M$-fold joint Gaussian
probability density function
\be p([A])=\dsf{1}{(2\pi)^M |C_A| }\exp\left(-\dsf{[A^*][A]}{[C_A]}
\right)\, \lab{lspeck2} \ee
for far-fields $A(x,y)$ at different points $x,y$. Here $[C_A]$ is a
Hermitian symmetric matrix with determinant $|C_A|$, whose elements
are given by $(C_A)_{i,j}=\langle A^*(x_i,y_i) A(x_j,y_j)\rangle$
for a set of far-fields $[A] \equiv \{A(x_1,y_1),A(x_2,y_2),\ldots
A(x_M,y_M)\}$ at $M$ points of the $x,y$ plane. Note that the
notation $\langle \cdots \rangle$ in the expressions for $\eta^2$
and $(C_A)_{i,j}$
and throughout below in the text means the disorder ensemble average.
Furthermore, one assumes that the indices $i,j$ at $(C_A)_{i,j}$ are taken
for adjacent spatial positions.

Expressions~\re{lspeck1} and~\re{lspeck2} are the result of the
central limit theorem of probability theory~\ci{Middleton}, which
claims the following: if complex random variables are the sum of
other independent complex random variables then, at the increase of
the number of second ones, the first ones are distributed according to
the Gaussian law. Applying the theorem for our case we have
far fields $A(x,y)$ as result of the interference of independent
object waves at all positions of $x,y$ plane. As we will see below, there are mainly two
physical conditions for the validity of the central limit
theorem. They are related to the physics of providing independence of the
object waves and to the method of their summation within
the interference process. We will now describe both of them in more detail.

A requirement for the object wave $a(\alpha,\beta)$ to be
independent leads to some limitations for its statistical
properties~\ci{Dainty}. First of all, formed after the reflection of
monochromatic light from the rough surface,  the individual wavelet
$a(\alpha,\beta)$ should be completely polarized. Second, the
first-order probability density of its phase should be uniform in
the interval $-\pi$ to $\pi$.  And at last, the object wave
$a(\alpha,\beta)$ should be quasi-homogeneous, which means that its
autocorrelation function $C_a$ consists of a slowly-varying
intensity $I_a$ envelope and a short-range normalized correlation
function $C_a'$:
\bea
C_a(\alpha_1 ,\beta_1 ;\alpha_2 ,\beta_2) \equiv \langle
a^*(\alpha_1 ,\beta_1) a(\alpha_2 ,\beta_2)\rangle
\nonumber \\
=I_a\left(
\dsf{\alpha_1+\alpha_2}{2},\dsf{\beta_1+\beta_2}{2}\right)
C_a'(\alpha_2-\alpha_1,\beta_2-\beta_1)\,.
\lab{lspeck3}
\eea
If we increase in this expression the range of variation of the
correlation part, i.e.~the correlation length of the object wave,
the changing range of the intensity becomes smaller. However, in
order to entirely satisfy the independence condition of the object
waves, their correlation length in Eq.~\re{lspeck3} should be as
short as possible, which means that $C_a'(\alpha_,\beta)$ has to be
delta correlated. The latter introduces some demands upon the
properties of the random light scatterer, which is called in the
literature as the diffusor. Typically, a diffusor is an optically
homogeneous transparent glass plate with no reflection centers for
light in the volume and a geometrically inhomogeneous distribution
of reflection centers with random heights on its surface. As
mentioned in the introductory section of the paper, the scattering
rough surface generates object waves in a plane $\alpha,\beta$,
which is closely situated at the surface, when the monochromatic
polarized incident light transmits through the plate.
Another realization of object waves is considered in
Refs.~\ci{Goodman1,Goodman2}, where the lateral monochromatic light
was directly incident on the rough surface. For our purpose to
calculate the  normalized speckle autocorrelation function, the
optical property of the medium, from which light falls on the
rough surface, is merely dropped from the consideration.
Assuming a Gaussian probability density of the surface height
$h(\alpha,\beta)$ with the autocorrelation function
$C_h(\alpha,\beta)$ and a variance $\eta_h^2$ and assuming also a
Gaussian probability density of object wave phases with a variance
$\eta_\phi^2$, Goodman derived the relation~\ci{Goodman1,Goodman2}
\be
C_a'(\alpha_,\beta)=\exp \left( -\eta_\phi^2
[1-C_h(\alpha,\beta)] \right)\,,
\lab{lspeck4}
\ee
where $\eta_\phi= 2\pi \eta_h /\lambda$. This function can be
approximated by a delta function $\delta (\alpha_,\beta)$ when
$\eta_\phi>1$ and thus $\eta_h>\lambda/2$ and the mean distance
between two inhomogeneous $h(\alpha,\beta)$  is larger than
$\lambda$. The delta functional autocorrelation of object waves
provides their independence from each other. On the other hand, the
Gaussian probability density of phases reduces to a uniform one for
$\eta_\phi>1$ supporting the second requirement for the object waves
outlined above. Therefore, the requirements for object waves
described in the previous paragraph can be experimentally realized
if the size of the surface inhomogeneities and the distance between
them are larger than the wave length of the light. As a next
implication of the presented analysis we can suppose that for the
outlined system parameters the object wave probability density
itself may have a circular Gaussian form for the real and imaginary
components of the wavelet $a(\alpha,\beta)$.

The next important step of the theory of Gaussian laser speckles is
the formation of far fields for a given set of object waves. It is
based on the Huygens-Fresnel principle of optics, which preserves
the individual wavelet picture, i.e., works in the limit of optics,
where deviations from geometrical optics are small. The interference
of the object waves yields an amplitude $A(x,y)$, which reads in the
above mentioned far field Fresnel approximation as follows
\be
A(x,y)=\int_{\cal A}
a(\alpha,\beta)\exp \left[ -\dsf{2\pi i}{\lambda z} (x\alpha +
y\beta)\right] d\alpha d\beta \,,
\lab{lspeck5}
\ee
where we have
omitted unimportant multipliers in front and inside of the integral.
This expression resembles, indeed, a Fourier
transformation and, thus, conserves the
principle that each object wave contributes individually to the
interference.

Using Eqs.~\re{lspeck3} and~\re{lspeck5} it is straight-forward to derive the
expression for the autocorrelation function
\be
C_A(x_1 ,y_1 ;x_2,y_2) = \langle A^*(x_1 ,y_1) A(x_2 ,y_2)\rangle \,.
\lab{lspeck6}
\ee
It turns out to be given by
\be
C_A(x_1 ,y_1 ;x_2 ,y_2) = I_A\left(
\dsf{x_1+x_2}{2},\dsf{y_1+y_2}{2}\right) C_A'(\Delta x,\Delta y)
\lab{lspeck7}
\ee
for $\Delta x=x_2-x_1$ and $\Delta y=y_2-y_1$ with
\be
I_A(x,y)=\int_{\cal A} C_a'(\alpha'',\beta'')\exp \left[
-\dsf{2\pi i}{\lambda z} (x\alpha'' + y\beta'')\right] d\alpha''
d\beta''
\lab{lspeck8}
\ee
and
\be
C_A'(x,y)=\int_{\cal A}
I_a(\alpha',\beta')\exp \left[ -\dsf{2\pi i}{\lambda z} (x\alpha' +
y\beta')\right] d\alpha' d\beta'
\lab{lspeck9}
\ee
for the novel variables $\alpha'=(\alpha_1+\alpha_2)/2$,
$\beta'=(\beta_1+\beta_2)/2$ and $\alpha''=\alpha_2-\alpha_1$,
$\beta''=\beta_2-\beta_1$. Eq.~\re{lspeck7} shows that the far
fields $A(x,y)$ are
quasi-homogeneous like the object waves $a(\alpha,\beta)$.
This is a direct result of the Fourier transformation
(\ref{lspeck5}). Another important result of this linear
transformation is the implicit proof of the above made supposition
that object waves are circularly Gaussian distributed. Indeed, only
Gaussian distributed object waves can contribute through the linear
mapping to Gaussian far fields. As we will see below the role of the
so far uninvestigated parameter, the aperture $\cal A$, will lead to
the formation of a correlation length of the correlation function
$C_A'(x,y)$.

The autocorrelation function $C_I(x,y)$ of the far-field intensity
$I(x,y)=|A(x,y)|^2$ can be calculated using the Wick theorem for
variables distributed according to the Gaussian law. A simple
calculation gives the expression
\be
C_I(x,y)=\langle I \rangle^2 \left[1+|C_A'(x,y)|^2\right]\,,
\lab{lspeck10}
\ee
where the
normalized autocorrelation function for the far-field $C_A'(x,y)$ is
defined as $C_A'(x,y)/C_A'(0,0)$, where $C_A'(0,0)=\langle I \rangle$.

As already mentioned above, if in Eq.~\re{lspeck3} the autocorrelation
function of object waves $C_a'(\alpha_,\beta)$ is delta correlated,
then the intensity function of these waves $I_a(\alpha,\beta)$ can
be approximated as a constant.
Assuming that the $\alpha,\beta$ plane is close to the
rough surface, one can write $|a(\alpha,\beta)|=\kappa
|P(\alpha,\beta)|$, where $P(\alpha,\beta)$ is the incident to the
glass plate light wave and $\kappa$ is the average reflectivity of
surface, for each position $\alpha,\beta$. Then the intensity of the
object waves at $\alpha,\beta$ is determined by the relation
$I_a(\alpha,\beta)=\kappa^2 |P(\alpha,\beta)|^2$. Therefore, the
expression for the normalized far-field autocorrelation function
reads
\be
C_A'(x,y)=\dsf{\int_{\cal A} |P(\alpha,\beta)|^2 \exp
\left[ -\dsf{2\pi i}{\lambda z} (x\alpha + y\beta)\right] d\alpha
d\beta}{\int_{\cal A} |P(\alpha,\beta)|^2 d\alpha d\beta}\,.
\lab{lspeck11}
\ee
\section{Speckle autocorrelation function for apertures in $1d$
to $3d$ dimensions} \label{sec3}

As already mentioned in the introductory section, the investigation
of a BEC in the laser speckle disorder has found much attention from
both a theoretical and an experimental point of view. In particular,
a variety of isotropic and anisotropic speckles have been the
subject of these works.
Motivated by this interest, we will describe in the present section
the derivation of the speckle autocorrelation function for different
apertures ranging from one to three dimensions by generalizing 
the appropriate expressions from the previous section to these dimensions.

\subsection{Real space}

Due to the analytic form of Eq.~\re{lspeck11}, we can take the
intensity of the incident wave $|P(\alpha,\beta)|^2$ to be unity
over the whole aperture region of the $\alpha,\beta$ plane. Writing
the function $|P|^2$ in the form $|P|_{d,{\cal A}}^2$, where $d$ is
the space of dimensionality and ${\cal A}$ is the form of the
aperture, we have the following expressions:
%
%
%
\be
|P(\alpha,\beta)|_{2d,{\rm rct}}^2={\rm rect}
\left( \dsf{\alpha}{L_{\alpha}}\right) {\rm rect} \left(
\dsf{\beta}{L_{\beta}}\right)
\lab{lspeck14}
\ee
for the $2d$ anisotropic rectangular aperture with sizes $L_{\alpha}$
and $L_{\beta}$, where the function ${\rm rect} (x)=1$ for $|x|\leq
1/2$ and zero otherwise;
retaining in Eqs.~\re{lspeck14} only the first ${\rm rect} (x)$
function and equating $L_{\alpha}=L$ one obtains the expression of
$|P(\alpha)|_{1d,{\rm inv}}^2$ for the $1d$ interval aperture of the
size $L$; the analytic form of $|P(\alpha,\beta)|_{2d,{\rm qdt}}^2$
for the $2d$ quadratic aperture of the size $L$ is obtained from
Eqs.~\re{lspeck14} if we specialize this equation according to
$L_{\alpha}=L_{\beta}=L$;
\be |P(\alpha,\beta)|_{2d,{\rm crc}}^2={\rm circ} \left(
\dsf{2 r}{D}\right)
\lab{lspeck15}
\ee
for the $2d$ isotropic circular
aperture with the diameter $D$ and $r=\sqrt{\alpha^2+\beta^2}$, where the
function ${\rm circ} (x)=1$ for $|x|\leq 1$ and zero otherwise;
%
%
%
\be
|P(\alpha,\beta,\gamma)|_{3d,{\rm rcpl}}^2={\rm rect}
\left( \dsf{\alpha}{L_{\alpha}}\right) {\rm rect} \left(
\dsf{\beta}{L_{\beta}}\right){\rm rect} \left(
\dsf{\gamma}{L_{\gamma}}\right)
\lab{lspeck17}
\ee
for the $3d$
anisotropic rectangular parallelepiped aperture of sizes
$L_{\alpha}$, $L_{\beta}$ and $L_{\gamma}$;
the expression $|P(\alpha,\beta,\gamma)|_{3d,{\rm cub}}^2$ of the
$3d$ cubic aperture of the size $L$ is obtained from
Eq.~\re{lspeck17} by setting $L_{\alpha}=L_{\beta}=L_{\gamma}=L$;
\be
|P(\alpha,\beta,\gamma)|_{3d,{\rm sph}}^2={\rm circ} \left( \dsf{2 r}{D}\right)
\lab{lspeck18}
\ee
for the $3d$ isotropic sphere aperture with the
diameter $D$ and $r=\sqrt{\alpha^2+\beta^2+\gamma^2}$;
\be
|P(r,\gamma)|_{3d,{\rm cyl}}^2={\rm circ} \left( \dsf{2 r}{D}\right){\rm rect} \left(
\dsf{\gamma}{L_{\gamma}}\right)
\lab{lspeck19}
\ee
for the $3d$ anisotropic cylinder aperture with the diameter of
circle $D$, $r=\sqrt{\alpha^2+\beta^2}$ and size $L_{\gamma}$ along
the $\gamma$ axis.
Substituting the expressions~\re{lspeck14}--\re{lspeck19} of the
$|P|_{d,{\cal A}}^2$ function in Eq.~\re{lspeck11} and calculating
the respective integrals, we obtain the corresponding expressions
for the correlation function $|C_A'|^2$:
%
%
%
%
%
\be |C_A'(\Delta x,\Delta y)|_{2d,{\rm rct}}^2= {\rm sinc}^2
\left( \dsf{L_{\alpha}\Delta x}{\lambda z}\right){\rm sinc}^2 \left(
\dsf{L_{\beta}\Delta y}{\lambda z}\right)
\lab{lspeck22}
\ee
where ${\rm sinc}(y)=\sin(\pi y)/(\pi y)$, for the $2d$ anisotropic
rectangular aperture with $z$ being the distance between object wave
and far field planes; retaining in this equation only first ${\rm
sinc}^2 (y)$ function, the dependence on $\Delta x$ and assuming
$L_{\alpha}=L$ one obtains the expression $|C_A'(\Delta x)|_{1d,
{\rm inv}}^2$ for the $1d$ interval aperture with $z$ being the
distance between object wave and far field intervals; the expression
$|C_A'(\Delta x,\Delta y)|_{2d,{\rm qdt}}^2$ for the $2d$ quadratic
aperture one can derive from Eq.~\re{lspeck22} for
$L_{\alpha}=L_{\beta}=L$;
\be
|C_A'(r)|_{2d,{\rm crc}}^2=\left|
2\dsf{J_1\left( \dsf{\pi Dr}{\lambda z}\right)}{\dsf{\pi Dr}{\lambda
z}} \right|^2\,,
\lab{lspeck23}
\ee
where $J_1(x)$ is a Bessel
function of the first kind and of the first order, for the $2d$
isotropic circular aperture with $r=\sqrt{(\Delta x)^2+(\Delta y)^2}$;
%
%
%
\begin{eqnarray}
&& \hspace*{0.5cm} |C_A'(\Delta x,\Delta y,\Delta z)|_{3d,{\rm rcpl}}^2= \nonumber\\
&&{\rm sinc}^2 \left( \dsf{L_{\alpha}\Delta x}{\lambda z}\right){\rm sinc}^2
\left( \dsf{L_{\beta}\Delta y}{\lambda z}\right){\rm sinc}^2 \left(
\dsf{L_{\gamma}\Delta z}{\lambda z}\right)
\lab{lspeck25}\end{eqnarray}
for the $3d$ anisotropic rectangular parallelepiped aperture with
$z$ as the distance between the object wave and the far field
volumes; the expression $|C_A'(\Delta x,\Delta y,\Delta z)|_{3d,{\rm
cub}}^2$ for the $3d$ cubic aperture is obtained from
Eq.~\re{lspeck25} by assuming
$L_{\alpha}=L_{\beta}=L_{\gamma}=L$;
\begin{eqnarray}
&&\hspace*{1cm}|C_A'(r)|_{3d,{\rm sph}}^2= \lab{lspeck26}  \\
&&\left| 3\left( \dsf{\lambda z}{\pi
Dr}\right)^3\left[ \sin \left( \dsf{\pi Dr}{\lambda z}\right)-\left(
\dsf{\pi Dr}{\lambda z}\right) \cos \left( \dsf{\pi Dr}{\lambda
z}\right)\right] \right|^2
\nonumber
\end{eqnarray}
for the $3d$
isotropic sphere aperture with $r=\sqrt{(\Delta x)^2+(\Delta
y)^2+(\Delta z)^2}$;
\be
|C_A'(r,\Delta z)|_{3d,{\rm cyl}}^2=\left|
2\dsf{J_1\left( \dsf{\pi Dr}{\lambda z}\right)}{\dsf{\pi Dr}{\lambda
z}} \right|^2 {\rm sinc}^2 \left( \dsf{L_{\gamma}\Delta z}{\lambda
z}\right)
\lab{lspeck27}
\ee
for the $3d$ anisotropic cylinder aperture with $r=\sqrt{(\Delta
x)^2+(\Delta y)^2}$.

Expressions of the autocorrelation function for a $2d$ quadratic
aperture
$|C_A'(\Delta x,\Delta y)|_{2d,{\rm qdt}}^2$ and for a
$2d$ circular aperture in Eq.~\re{lspeck23} are derived by Goodman
in Refs.~\ci{Goodman1,Goodman2}. As can be seen from the formulas
for other cases of the aperture, they are closely related to both of
these Goodman cases of the aperture. However, the derivation of the $3d$
isotropic sphere autocorrelation function (\ref{lspeck26}), which is
a result of the present paper, required some additional effort.

The analytical forms of the autocorrelation functions $|C_A'|^2$ are
similar in every spatial direction. They have one central maximum and a set
of side maxima of decaying height, which are separated from each other by
zeros. As was pointed out in the introductory
section, it is obvious that
these forms can not be fitted by a Gaussian. The
argument of the autocorrelation function, which corresponds to its first zero, provides the
correlation length of the disorder, i.e.~the average size of the speckle grain,
for the appropriate spatial direction. Denoting it by
$\delta x$ we have, for instance, for $1d$ speckle
\be
\delta x=\dsf{\lambda z}{L}\,.
\lab{lspeck27a}
\ee
The main interest of the present paper is the $3d$ spherical
aperture of Eq.~(\ref{lspeck26})
since we will carry out the calculation of BEC properties
for this particular case of laser speckles. Numerically solving the equation
$\sin(x)-x \cos(x)=0$ we find first its solution to be at $x_c=4.493$, thus
the disorder correlation length is given by $r_c=1.4302~\lambda z/D$.

In order
to establish a physical meaning of $\delta x$ we introduce the
"wave number" $k_{\rm eff}$, which is related to the vector
$\alpha,\beta$ in the above Fourier transform formulas, by the relation
$k_{\rm eff}=2\pi x/(\lambda z)$. If we substitute in it $\delta x$ from
Eq.~\re{lspeck27a} then we obtain $k_{\rm eff}=2\pi/L$. For a circular and
a spherical aperture the "wave number" is $k_{\rm eff}=2\pi/D$. However, the
sense of $k_{\rm eff}$ is in an uncertainty of the wave vector when the
problem of wave propagation is solved in the restricted area. It is
well known that in this area the wave vector is determined within the
resolution $k_{\rm eff}$. Therefore, we can say that the origin of a
speckle grain with a correlation length $\delta x$ as its size represents the
spatial uncertainty in the determination of far fields, which is introduced by the
finite size of the aperture.

\subsection{Fourier space}

For many applications the Fourier transform of the far-field intensity
autocorrelation function, or the power spectral density, of the
speckle is of considerable interest. In the literature on laser
speckle theory~\ci{Goodman1,Goodman2} it is defined according to
\be
C_I({\bf k})=\int C_I({\bf x})e^{-i2\pi {\bf k}{\bf x}} d^d x\,.
\lab{lspeck28}
\ee
Substituting
in it Eq.~\re{lspeck10} for $C_I({\bf x})$ one obtains
\be
C_I({\bf k})= \langle I \rangle^2 \left[\delta ({\bf k})+|C_A'({\bf k})|^2\right]\,.
\lab{lspeck29}
\ee
In the perturbative considerations of BEC in the speckle potential
the Fourier transform $|C_A'({\bf k})|^2$ plays the central role. It
has the following expressions for the real space autocorrelation
functions taken from Eqs.~\re{lspeck22}--\re{lspeck27}:
%
%
%
%
\be
|C_A'({\bf k})|_{2d,{\rm rct}}^2= \dsf{(\lambda z)^2}{L_{\alpha} L_{\beta}} {\rm tri}
\left( \dsf{k_x\lambda z}{L_{\alpha}}\right){\rm tri} \left(
\dsf{k_y\lambda z}{L_{\beta}}\right)
\lab{lspeck32}
\ee
where the triangle function is defined as ${\rm tri}(x)=1-|x|$ for
$|x|\leq 1$ and zero otherwise, for the $2d$ anisotropic rectangular
aperture; the expression $|C_A'({\bf k})|_{1d,{\rm inv}}^2$ for the
$1d$ interval aperture one can get from Eq.~\re{lspeck32} by
assuming $L_{\alpha}= L_{\beta}=L$, $k_x=k_y=k$ and taking the
square root of its right-hand side; the expression
$|C_A'({\bf k})|_{2d,{\rm qdt}}^2$ for the $2d$ quadratic aperture is
obtained from Eq.~\re{lspeck32} with the assumption $L_{\alpha}=
L_{\beta}=L$;
\begin{eqnarray}
&& |C_A'({\bf k})|_{2d,{\rm crc}}^2=  2\left(\dsf{2\lambda z}{\pi D}\right)^2
\lab{lspeck33} \\
&&\times \left[ \cos^{-1}\left(
\dsf{k\lambda z}{D}\right)-\dsf{k\lambda z}{D}\sqrt{1-\left(
\dsf{k\lambda z}{D}\right)^2}\,\right] \nonumber
\end{eqnarray}
for the
$2d$ isotropic circular aperture with $k=\sqrt{k_x^2+k_y^2}$;
%
%
\begin{eqnarray}
&&|C_A'({\bf k})|_{3d,{\rm rcpl}}^2=
\dsf{(\lambda z)^3}{L_{\alpha} L_{\beta} L_{\gamma}}
\lab{lspeck35} \\
&&\times {\rm tri} \left( \dsf{k_x\lambda z}{L_{\alpha}}\right) {\rm tri} \left(
\dsf{k_y\lambda z}{L_{\beta}}\right) {\rm tri} \left( \dsf{k_z\lambda
z}{L_{\gamma}}\right)\nonumber
\end{eqnarray}
for the $3d$
anisotropic rectangular parallelepiped aperture;
the expression $|C_A'({\bf k})|_{3d,{\rm cub}}^2$ for the $3d$ cubic
aperture is obtained from Eq.~\re{lspeck35} by specializing $L_{\alpha}=
L_{\beta}= L_{\gamma}=L$;
\be
|C_A'({\bf k})|_{3d,{\rm sph}}^2=\dsf{3}{\pi}\left(\dsf{2\lambda
z}{4D}\right)^3(b^3-12b+16)
\lab{lspeck36}
\ee
for the $3d$
isotropic sphere aperture with $b=2k\lambda z/D$ and
$k=\sqrt{k_x^2+k_y^2+k_z^2}$;
\begin{eqnarray}
&&|C_A'({\bf k})|_{3d,{\rm cyl}}^2= \left(\dsf{\lambda z}{L_{\gamma}}\right) {\rm tri} \left( \dsf{k_z\lambda
z}{L_{\gamma}}\right) \lab{lspeck37} \\
&&\times 2\left(\dsf{2\lambda z}{\pi D}\right)^2 \left[ \cos^{-1}\left(
\dsf{k\lambda z}{D}\right)-\dsf{k\lambda z}{D}\sqrt{1-\left(
\dsf{k\lambda z}{D}\right)^2}\,\right]
\nonumber
\end{eqnarray}
for the $3d$
anisotropic cylinder aperture with $k=\sqrt{k_x^2+k_y^2}$.

The equation for the quadratic aperture $|C_A'({\bf k})|_{2d,{\rm
qdt}}^2$ and Eq.~\re{lspeck33} have been derived by Goodman in
Refs.~\ci{Goodman1,Goodman2}. Other expressions of $|C_A'({\bf
k})|^2$, except for the $3d$ isotropic sphere aperture case, can be
obtained by using these formulas. Eq.~\re{lspeck36} is a result of
this paper. In all our formulas for the anisotropic aperture we have
assumed that the size deviation of the aperture with respect to its
average isotropic size is essentially less than the distance $z$.

As is seen from the formulas of $|C_A'({\bf k})|^2$ expressed through the
triangle function their value becomes zero when their argument  is
unity. For the $2d$ circle and the $3d$ sphere apertures $|C_A'({\bf k})|^2$
is zero for $k\lambda z/D=1$. Hence, the wave vector of the Fourier
transform autocorrelation function only varies in a finite interval
from zero, in contrast to the case for a Gaussian
function. This fact is another reason why the speckle autocorrelation
function can not be approximated by a Gaussian form.

It is worth to discuss the expression $|C_A'(r)|^2={\rm sinc}^2(k_L
r)$ with $k_L= D/\lambda z$ for the autocorrelation function used
in Ref.~\ci{Kuhn} for the $3d$ isotropic aperture. It is similar to our correlation function
$|C_A'(\Delta x)|_{1d,{\rm inv}}^2$ for the $1d$ interval aperture.
The authors of Ref.~\ci{Kuhn} claim that this expression is valid
for $z \sim (\alpha^2+\beta^2)_{\rm max}/\lambda$, which is outside
of the far-field limit. However, that limit destroys the
fundamentals of the Gaussian speckle theory as they are described in
Sec.~\ref{sec2}. Therefore, it is unclear whether $|C_A'(r)|^2$ of
Ref.~\ci{Kuhn} is related to laser speckles or not.

Furthermore, we discuss the definition of the speckle correlation
length to be the width at the half value of the maximum of
$|C_A'(r)|^2$ for $r=0$, when the last one is approximated by a
Gaussian function.
Probably, this definition was introduced first by Modugno in
Ref.~\ci{Modugno}, when he considered $|C_A'(\Delta x)|_{1d,{\rm
inv}}^2$. It was found in Ref.~\ci{Modugno} that the correlation
length is given by $\delta x=0.88\lambda z/L$, while from
Eq.~\re{lspeck27a} the exact value turns out to be $\delta x=\lambda
z/L$. It is interesting that the Gaussian $|C_A'(r)|^2$ has been
obtained in the numerical simulation of the $3d$ isotropic laser
speckle in Refs.~\ci{Pilati1,Pilati2} which should be compared with
the exact $|C_A'(r)|_{3d,{\rm sph}}^2$ in Eq.~\re{lspeck26}, with
the correlation length $r_c=1.1\lambda z/D$, however, the exact one
is $r_c=1.4302 \lambda z/D$, see the discussion after
Eq.~\re{lspeck27a}. It seems that we can explain the reason why the
authors of Refs.~\ci{Pilati1,Pilati2} obtained the Gaussian form of
$|C_A'(r)|^2$. They used the speckle simulation method proposed by
Huntley in Ref.~\ci{Huntley} which we briefly review for the $2d$
case. Let us consider to this end two square planes $\alpha,\beta$
and $x,y$ with the same size $L$. According to the Huntley method
one uses Eq.~\re{lspeck5} in order to perform a double Fourier
transformation. In the first inverse Fourier transformation the
complex object waves $a(\alpha,\beta)$ on the mesh points in the
$\alpha,\beta$ plane are simulated through the given Gaussian
distributed complex random waves $A(x,y)$ on the mesh points in the
$x,y$ plane. Afterwards, one cuts by a circle with radius $D/2$ the
$\alpha,\beta$ region of the obtained $a(\alpha,\beta)$ such that it
vanishes outside of this region. In the second direct Fourier
transformation the derived complex waves $a(\alpha,\beta)$ form the
final complex far-fields $A(x,y)$. Huntley has investigated in
Ref.~\ci{Huntley} only the first-order statistical property of the
simulated pattern, i.e.~the probability density of the intensity,
and showed that it corresponds to the theoretical laser speckles of
Ref.~\ci{Goodman1}. However, the proposed simulation method can
drastically deviate  in the second-order statistical property of a
speckle, i.e.~its autocorrelation function, from the theoretical
one.

Indeed, in accordance with the theory of a speckle autocorrelation
function as presented in this section, after the first Fourier mapping the
object waves $a(\alpha,\beta)$ acquire a correlation with the
correlation length $\delta \alpha=\delta \beta=\lambda z/L$, where
$L$ is size of the square $x,y$ plane. More precisely, now the
function $C_a'(\alpha,\beta)$ is not delta correlated. However,
according to Eq.~\re{lspeck3}, the broadening of the
$C_a'(\alpha,\beta)$ function reduces to a changing of a constant
character of the $I_a(\alpha,\beta)$ function to one of varying in space in the
$\alpha_,\beta$ plane. Substituting this function of
$I_a(\alpha,\beta)$ in Eq.~\re{lspeck9} and integrating over
$\alpha$ and $\beta$ gives the function $C_A'(x,y)$ which may
qualitatively be different from the one discussed in this section.

A simulation method, which is consistent with the above laser speckle theory,
is described in the book of Goodman~\ci{Goodman2}. There are other
numerical methods in Refs.~\ci{Makse1,Makse2}, in which the exact form of the real
space autocorrelation function is used to generate the speckle
pattern. In particular, one of such methods was exploited for the
simulation of $1d$ speckle in Ref.~\ci{Sucu}.

\section{Experimental realization of $3d$ isotropic speckle}
\label{sec4}

As was already mentioned in the introductory section, we consider here a true
$3d$ speckle, not the quasi-three dimensional one consisting of a
transverse $2d$ speckle with a longitudinal depth in the
autocorrelation function as described in details in
Ref.~\ci{Leushacke} and section 4.4.3 of the Goodman
book~\ci{Goodman2} and applied in many experiments. At a first
glance, it seems exotic and unrealistic to experimentally realize
such a $3d$ volume speckle pattern. However, in the present section
we will describe the physical principle how it can be generated.

In the  typical $2d$ geometry of the experimental realization of a
speckle a lens, which collects the incident light, is installed close to the
glass plate such that its focal plane coincides with the far-field
plane~\ci{Clement2}. This idea of a speckle formation in the focal
plane can be generalized to a full $3d$ geometry, when the speckle is formed in
the focal point, i.e.~the focus, of an empty ellipsoidal optic cavity according to
the scheme displayed in Fig.~1.
\begin{figure}[t]
\begin{center}
\includegraphics[scale=0.5]{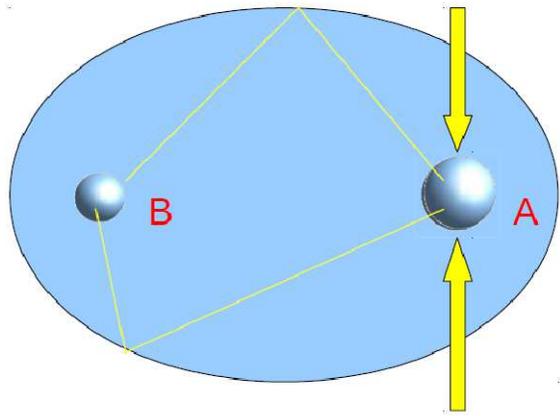}
\end{center}
\caption{Cross section of the ellipsoidal reflective cavity with spheres
$A$ and $B$ in its focuses. Focus $A$ contains a small size absolute
spherical light reflector in the center and volume optic
inhomogeneities. Incident laser beams (thick yellow arrows), after
reflection from the reflector, scatter additionally from inhomogeneities
producing individual wavelets (thin yellow lines), which are
collected in focus $B$, where a BEC is deposited.}
\lab{fig1}
\end{figure}
Let us consider that cavity, whose inside surface
reflects absolutely the light emitted from one of its focus (point $A$) and
collects it at the second focus (point $B$). Two laser beams (thick
yellow arrows) are incident through holes in the cavity surface into
the small metallic sphere, i.e.~the reflector, with absolute light
reflection, located in the center of the glass sphere $A$. It is assumed
that laser beams cover the entire surface of this reflector and a BEC
is deposited in the sphere $B$, which is located at the second focus
of the ellipsoid.

The glass sphere $A$ additionally contains the located randomly
light scattering centers, for instance, absolutely light reflective
metallic polyhedrons with a random average size of each facet. The
theory how to derive the $2d$ object wave autocorrelation function
$C_a'(\alpha,\beta)$, described in Refs.~\ci{Goodman1,Goodman2}, can
be easily generalized to the derivation of
$C_a'(\alpha,\beta,\gamma)$ for such a $3d$ case with the same
expression~\re{lspeck4}. However, now this expression is a function
of $3d$ other quantities. The condition, at which
$C_a'(\alpha,\beta,\gamma)$ becomes delta correlated and the object
waves are independent, is the same as for $C_a'(\alpha,\beta)$.
Therefore, if the mean distance between these light scattering
centers and the average size of polyhedrons are larger than the
light wavelength, then $C_a'(\alpha,\beta,\gamma)$ will be delta
correlated. On the other hand, each sphere with radius
$(\alpha^2+\beta^2+\gamma^2)^{1/2}$ and with the same center as the
sphere $A$ will be the object wave volume, whereas for comparison
for $2d$ we had a $\alpha,\beta$ object wave plane.

Incident laser beams, after reflection from the reflector, scatter
additionally from scattering centers and produce individual and
independent wavelets, the object waves $a(\alpha,\beta,\gamma)$, indicated via thin
yellow lines in Fig.~\ref{fig1}, which are collected in the sphere $B$, where a BEC
is deposited. For the presented geometry the far-field condition is
satisfied, since a distance $z$ between the object wave and the far-field
volumes, i.e.~the length of each wavelet trajectory between two focuses
of the ellipsoid, is larger than the size of the object wave sphere
$A$.

The described scheme can be generalized for the experimental
realization of any $3d$ anisotropic speckle. To this end one only needs to
change the form of the spherical aperture $A$, which contains
the glass and the light scattering centers, into a suitable one listed in
the previous section. The spherical form of the
metallic reflector retains unchanged.

At the end of this section, it is worthwhile to discuss the possible
realization of a $3d$ volume speckle pattern using $2d$ plane
speckles. Such a scenario presumes a $3d$ speckle as a result of the
sum or, more clearly, as a linear interference of two and more $2d$
speckles. While theoretically this scenario is discussed by Pilati
{\it at} {\it al.} in Ref.~\ci{Pilati2}, the experiment, in which
two perpendicular $2d$ speckle planes form a $3d$ speckle pattern,
was realized in Ref.~\ci{Jendrzejewski} by Jendrzejewski {\it et}
{\it al.} Instantly the question arises whether the random pattern
realized in such a way belongs to the class of speckles or not. In
spite of an additional theoretical analysis, which is required to
answer that question in detail, the following argument shows that
the possible conclusion is negative. Indeed, according to the
fundamentals of the laser speckle theory of Goodman,
Refs.~\ci{Goodman1,Goodman2}, and Dainty, Ref.~\ci{Dainty}, see also
Secs.~\ref{sec2} and~\ref{sec3} of this paper, the correlated
speckle pattern in any dimension,
except the one described in Ref.~\ci{Leushacke} and its analogue for
$1d$ (see next paragraph), is a result of the Fourier transform over
the restricted aperture object wave region in the same dimension.
This means that a $3d$ volume speckle can be obtained only by a $3d$
object wave volume. Physically it means that the single connected
spatial domain of each $3d$ speckle grain, which is a result of $3d$
correlations, can not be obtained by a linear combination of randomly
sized and independent $2d$ speckle grains.
By that reason, the true $3d$ speckle cannot be obtained even by a
combination of quasi-three dimensional speckles,  which we
discussed at the beginning of this section.

\section{BEC depletion and sound velocity in weak $3d$ isotropic speckle}
\label{sec5}

The interaction potential of light with an atom at position ${\bf r}$ is determined by the far-field intensity $I({\bf r})=
|A({\bf r})^2|$ and has the form $V({\bf r})=t I({\bf r})$, see for instance
Refs.~\ci{Clement2,Kuhn}, where the constant $t$ is a function of
the atomic and light characteristics. At the derivation of
$V({\bf r})$ it was assumed that the incident laser wave does not induce an
atomic electron interlevel transition, but merely deforms the atomic
ground state.

It is convenient to define the interaction potential as $V({\bf r})=V_0+\Delta V({\bf r})$,
where $\Delta V({\bf r})=V({\bf r})-V_0$ and
$V_0=\langle I \rangle$. Using the obvious property $\langle \Delta V({\bf r})\rangle =0$, a simple
calculation shows that
\be \langle V({\bf r}')V({\bf r}'+{\bf r}) \rangle =
V_0^2 \left[ 1+\dsf{\langle \Delta V({\bf r}')
\Delta V({\bf r}'+{\bf r}) \rangle}{V_0^2}\right]
\lab{lspeck38}
\ee
and, therefore, we have the following relationships between the laser
speckle autocorrelation and the disorder potential correlation
functions:
\begin{eqnarray}
|C_I({\bf r})|^2&=& \langle V({\bf r}') V({\bf r}'+{\bf r}) \rangle \, , \nonumber \\
|C_A'({\bf r})|^2 &=&\dsf{\langle \Delta V({\bf r}') \Delta V({\bf r}'+{\bf r}) \rangle}{V_0^2} \, .
\lab{lspeck39}
\end{eqnarray}
Our interest is a Bose gas with a contact interaction. Taking
into account that, according to the novel definition of $V({\bf r})$, the
chemical potential for the ground state of BEC will be renormalized according to
$\mu \rightarrow \mu-V_0$, the Gross-Pitaevskii equation (GPE) reads
\be
\left[-\dsf{\hbar^2}{2m}{\mbox{\boldmath $\nabla$}}^2 + \Delta V({\bf r})
+g|\Psi({\bf r})|^2 - \mu \right] \Psi({\bf r})=0 \,.
\lab{lspeck40}
\ee
Here $g=4\pi \hbar^2 a/m$ denotes the strength of the contact interaction
with the scattering length $a$.

Under the assumption that the disorder potential is weak, one can
expand the solution
\be
\Psi({\bf r})=\psi_0+\psi_1({\bf r})+\psi_2({\bf r})+\cdots
\lab{lspeck41}
\ee
and solve the GPE (\ref{lspeck40}) perturbatively in the respective order of
$\Delta V({\bf r})$~\ci{Krumnow}. For the ground state all functions
of the expansion as well as $\Psi({\bf r})$ are real. In this way the problem is
reduced to find the total particle density $n=\langle \Psi({\bf r})^2\rangle$
and the condensate density  $n_0=\langle \Psi({\bf r}) \rangle^2$. In
particular, the lowest order expression for
the condensate depletion reads
\be
n-n_0=n_0\, \int \, \dsf{d^3 k}{(2\pi)^3} \dsf{R({\bf k})}{[\hbar^2 {\bf k}^2/2m + 2ng]^2} +\cdots\,,
\lab{lspeck42}
\ee
where we introduced the literature notation $R({\bf k})= R|C_A'({\bf k})|^2$ with
$R=V_0^2$.

In order
to further apply our formula Eq.~\re{lspeck36} for the $3d$
isotropic autocorrelation function  $|C_A'({\bf k})|_{3d,{\rm sph}}^2$, one
needs to make a remark. According to the definition in Eq.~\re{lspeck28},
the Fourier transforms of autocorrelation functions carry a physical dimension.
In particular, the correlation function
$|C_A'({\bf k})|_{3d,{\rm sph}}^2$, calculated with Eq.~\re{lspeck28}, is proportional to the inverse volume of the
$3d$ isotropic aperture $3/(4\pi)(2/D)^3$ times $(\lambda z)^3$.
If we introduce the correlation length as $\sigma=\lambda z/D$, then
the proportionality factor is $3(2\sigma)^3/(4\pi)$. In the following, we assume that
$|C_A'({\bf k})|_{3d,{\rm sph}}^2$ is already normalized by that factor.

It is convenient to introduce also the BEC coherence length according
to $\xi=[\hbar^2/(2mng)]^{1/2}=1/\sqrt{8\pi na}$.
Substituting the normalized correlation function $R|C_A'({\bf k})|_{3d,{\rm sph}}^2$
from Eq.~\re{lspeck36} in Eq.~\re{lspeck42} and performing the
integration, we get the expression $n-n_0=n_{\rm HM}f(\sigma/\xi)$,
where the depletion $n_{\rm HM}=[m^2R/(8\pi^{3/2}
\hbar^4)]\sqrt{n/a}$ was obtained by Huang and Meng in Ref.~\ci{Huang} (see also Ref.~\cite{Falco}) for
delta correlated disorder $R({\bf r})$ and the condensate depletion function is defined via
\begin{eqnarray}
f\left(\dsf{\sigma}{\xi}\right)&=&\dsf{1}{\sqrt{2}\pi}\dsf{\sigma}{\xi}
\left[4-\left(\dsf{8\sigma^2}{\xi^2}+6\right) \ln
\left(1+\dsf{\xi^2}{2\sigma^2}\right)\right. \nonumber \\
&& \left. + \dsf{4}{\sqrt{2}}\dsf{\xi}{\sigma} {\rm arctan}
\left(\dsf{\xi}{\sqrt{2}\sigma}\right)\right]\, .
\lab{lspeck43}
\end{eqnarray}
The function $f(\sigma/\xi)$, which is depicted in Fig.~\ref{fig2},
has the following asymptotics for small $\sigma/\xi$
\be
f\left(\dsf{\sigma}{\xi}\right)\approx
1-\dsf{14\sqrt{2}}{3\pi}\left(\dsf{\sigma}{\xi}\right)^3
-\dsf{18\sqrt{2}}{5\pi}\left(\dsf{\sigma}{\xi}\right)^5+\cdots
\lab{lspeck44}
\ee
and, correspondingly, for large $\sigma/\xi$
\be
f\left(\dsf{\sigma}{\xi}\right)\approx
\dsf{1}{2^{5/2}\pi}\left[\dsf{1}{3}
\left(\dsf{\xi}{\sigma}\right)^3-\dsf{1}{10}
\left(\dsf{\xi}{\sigma}\right)^5 \right]+\cdots
\lab{lspeck45}
\ee
Introducing the appropriate correlation length for each aperture,
as described in Sec.~\ref{sec3}, one can show that, when this
correlation length tends to zero, then the corresponding correlation
function $|C_A'({\bf r})|^2$ tends to the delta function. The same
behavior has our function $|C_A'({\bf r})|_{3d,{\rm sph}}^2$ in the limit $\sigma
\rightarrow 0$. Therefore, we should reproduce the Huang and Meng
result $n_{\rm HM}$ for the condensate depletion in this limit.
Indeed, when $\sigma/\xi \rightarrow 0$ we read off from Eq.~\re{lspeck44} that one
obtains $f(\sigma/\xi)\rightarrow 1$.

\begin{figure}[t]
\begin{center}
\includegraphics[width=8cm,scale=1]{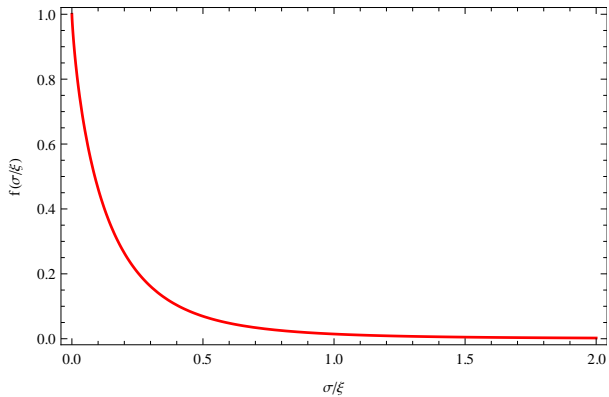}
\end{center}
\caption{Condensate depletion function $f(\sigma/\xi)$ from Eq.~(\ref{lspeck43}).}
\lab{fig2}
\end{figure}

For the $3d$ isotropic Bose gas with contact interaction the
normalfluid density $n_N$ is determined by the equation
$n_N=4(n-n_0)/3$ (see Sec.~\ref{sec6} below and Ref.~\ci{Krumnow} as
well as the references therein), from which $n_N$ is proportional to
the function of $f(\sigma/\xi)$.

In Ref.~\ci{Krumnow} the
sound velocity of a dipolar BEC in a weak external disorder potential is calculated
within a hydrodynamic approach. To this end a general derivation was performed
which is applicable for an arbitrary interaction potential.
For an isotropic $3d$ system with contact
interaction it has the form:
\begin{eqnarray}
\dsf{c}{c_0}&=&1+ \, \int \,
\dsf{d^3 k}{(2\pi)^3} \dsf{R({\bf k})}{(\hbar^2  {\bf k}^2/2m + 2ng)^2} \nonumber \\
&& \times \left\{ \dsf{\hbar^2 {\bf k}^2/2m}{(\hbar^2 {\bf k}^2/2m + 2ng)}-(\hat{\bf q}\hat{\bf k})^2 \right\} +\cdots\,,
\lab{lspeck46}
\end{eqnarray}
where $c_0=(ng/m)^{1/2}$ is the sound velocity in a system without
disorder and the scalar product between the sound direction $\hat{\bf q}$ and the direction of
wave propagation $\hat {\bf k}$ has the form $\hat{\bf q}\hat{\bf k}=\cos \vartheta$ for an
isotropic system.

Calculating the integral in Eq.~\re{lspeck46}, we obtain
$c/c_0=1+n_{\rm HM}s(\sigma/\xi)/(2n)$, where the sound velocity function reads
\begin{eqnarray}
s\left(\dsf{\sigma}{\xi}\right)&=&\dsf{2^{3/2}}{\pi}\dsf{\sigma}{\xi}
\left[\dsf{14}{3}-\left(\dsf{28\sigma^2}{3\xi^2}+4\right) {\rm ln}
\left(1+\dsf{\xi^2}{2\sigma^2}\right)\right. \nonumber \\
&& + \left. \dsf{5}{3\sqrt{2}}\dsf{\xi}{\sigma} \arctan
\left(\dsf{\xi}{\sqrt{2}\sigma}\right)\right]
\lab{lspeck47}
\end{eqnarray}
It is depicted in Fig.~\ref{fig3} and
has the following asymptotics for small $\sigma/\xi$
\be
s\left(\dsf{\sigma}{\xi}\right)\approx
\dsf{5}{3}+\dsf{2^{3/2} 3}{\pi}\left(\dsf{\sigma}{\xi}\right)
-\dsf{2^{3/2} 62}{9\pi}\left(\dsf{\sigma}{\xi}\right)^3+\cdots
\lab{lspeck48}
\ee
and for large $\sigma/\xi$
\be
s\left(\dsf{\sigma}{\xi}\right)\approx
\dsf{2^{1/2}}{\pi}\left[-\dsf{7}{3}
\left(\dsf{\xi}{\sigma}\right)+\dsf{13}{18}
\left(\dsf{\xi}{\sigma}\right)^3 \right]+\cdots \, ,
\lab{lspeck49}
\ee
respectively.

\begin{figure}
\begin{center}
\includegraphics[width=8cm,scale=1]{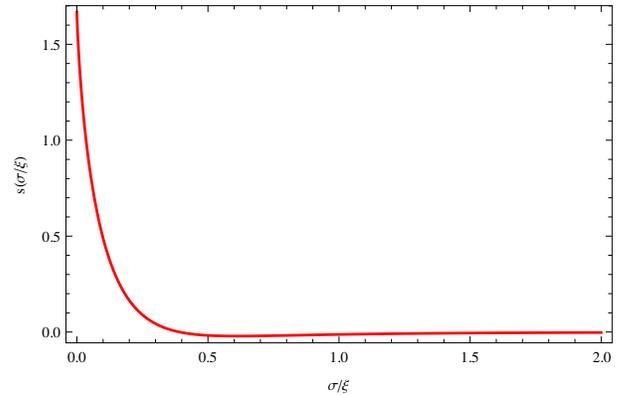}
\end{center}
\caption{Sound velocity function $s(\sigma/\xi)$ from Eq.~(\ref{lspeck47}).}
\lab{fig3}
\end{figure}

Again, when the correlation length $\sigma \rightarrow 0$ and thus
the correlation function $|C_A'({\bf r})|_{3d,{\rm sph}}^2$ is delta
correlated, we reproduce the result $s(\sigma/\xi)\approx 5/3$ of
Ref.~\ci{Giorgini}, obtained for delta correlated $R({\bf r})$.

As shown in this section, the finite range of integration for the
vector $\bf k$, when a Fourier transform of a speckle correlation
function is being applied, essentially simplifies the analytic
calculation of the BEC properties. This is an essential advantage of
applying the laser speckle theory to the BEC investigation. Conversely,
due to the infinite limit of integration on $\bf k$,
the Gaussian disorder correlation function, which is often used in
the literature, introduces some difficulties in its application to
the BEC theory.

\section{Landau derivation of normalfluid density}
\label{sec6}

Let $K_0$ be a reference frame, and $K$ a second frame with relative
velocity $- {\bf v}$ with respect to $K_0$. According to the
Galilean transformation in classical mechanics, the energy $E_0$ of
a system in the frame $K_0$ and its energy $E$ in the frame $K$ are
related to each other by:
\be E=E_0-{\bf P}_0 {\bf v} +  \frac{M}{2} {\bf v}^2 \,,
\lab{lspeck50} \ee
where ${\bf P}_0$ and $M$ are the total momentum and the mass of the
system, respectively.

Following to Refs.~\ci{Landau1a,Landau1b} let us assume that, at
temperature $T=0$, the condensate is in rest, i.e., in the frame
$K_0$, and its energy is $E_0=0$ with momentum ${\bf P}_0=0$. If one
quasiparticle with mass $m$ appears in the condensate with energy
$\varepsilon ({\bf p})$, where ${\bf p}$ is a momentum of the
quasiparticle, then in the frame $K_0$ energy and momentum now
become $E_0=\varepsilon ({\bf p})$ and ${\bf P}_0={\bf p}$. Hence,
from Eq.~\re{lspeck50} the energy $E$ in frame $K$ will be
$E=\varepsilon ({\bf p})-{\bf p}{\bf v}+M {\bf v}^2/2$ and the
energy of the quasiparticle in frame $K$ after a Galilean
transformation has a form $\varepsilon ({\bf p})-{\bf p}{\bf v}$.

According to the Landau two-fluid theory~\ci{Landau1a,Landau1b} of
liquid helium II, a gas of quasiparticles, for instance phonons,
constitutes the normalfluid density at low temperatures. For $T=0$
no quasiparticles exist, thus the helium is entirely superfluid. If
a gas of quasiparticles appears in the system for finite but low
temperatures, which has zero center mass velocity in the frame $K_0$
and moves with constant velocity $-{\bf v}$ with respect to the
frame $K$,
in which the helium liquid is in the rest, then the total momentum
of the gas per volume in the frame $K$ is given by
\be
\dsf{\bf P}{V}= \int {\bf p} \, N(\varepsilon ({\bf p})-{\bf p}{\bf v}) \dsf{d^3
p}{(2\pi \hbar)^3}\,,
\lab{lspeck51}
\ee
where $N(\varepsilon ({\bf p}))$ is the average occupation number of
states by phonons with energy $\varepsilon ({\bf p})$.
Eq.~\re{lspeck51} describes the thermodynamic property of a gas of
phonons. However, it can be generalized to our BEC system in the
external disorder potential at $T=0$, if we assume that, after
scattering with the disorder, particles of the condensate become the
quasiparticles of the normalfluid density.
It is clear that it occurs when the disorder is attached to the
frame $K_0$. To this end we replace in Eq.~\re{lspeck51} the
thermodynamic quantity $N(\varepsilon ({\bf p})-{\bf p}{\bf v})$ by
the quantum one $|\Psi ({\bf p}-m {\bf v})|^2$, where the wave
function is a solution of the GPE with disorder and written in
momentum representation. After that  we average both sides of
Eq.~\re{lspeck51} over the disorder ensemble. The obtained mean
square of the modulo of the wave function is now homogeneous in
space, so it can be expressed in terms of the energy of a
quasiparticle, as the Hamiltonian is commutative with the momentum
operator and thus the eigenfunction of the latter can be taken as
the eigenfunction of the former~\ci{Landau2}. Recalling that the
expression for the total density is  $n=\langle \Psi^2 \rangle$, we
obtain
\be
\dsf{\langle {\bf P} \rangle }{V}= \int
{\bf p} \, n(\varepsilon ({\bf p})-{\bf p}{\bf v}) \dsf{d^3 p}{(2\pi
\hbar)^3}\,.
\lab{lspeck52}
\ee

In order to derive the expression for the normalfluid density we
expand the integrand of Eq.~\re{lspeck52} in power of  ${\bf p}{\bf
v}$ and, in the limit ${\bf v} \rightarrow  {\bf 0}$, retain only
its first two terms. After integrating over the directions of the
vector ${\bf p}$ the zeroth order term of this expansion disappears.
Thus one obtains
\be \dsf{\langle{\bf P}\rangle}{V}= -\int {\bf p} \,({\bf p}{\bf v})
\dsf{d n(\varepsilon ({\bf p}))}{d\varepsilon ({\bf p})} \dsf{d^3
p}{(2\pi \hbar)^3}\,. \lab{lspeck53} \ee
This expression is the main result of the normalfluid density Landau
theory, when the two replacements $\langle {\bf P}\rangle$ by ${\bf
P}$ and $n(\varepsilon ({\bf p}))$ by $N(\varepsilon ({\bf p}))$ are
performed.

Taking into account that ${\bf p} ({\bf p}{\bf v})= p_z^2 {\bf v}$,
the expression for the normalfluid density reduces to
\be \rho_n= -\int p_z^2 \, \dsf{d n(\varepsilon ({\bf
p}))}{d\varepsilon ({\bf p})} \dsf{d^3 p}{(2\pi \hbar)^3}\,.
\lab{lspeck54} \ee
From Eq.~\re{lspeck42} we have the expression of
the total density Fourier transform
\be n(\varepsilon ({\bf p}))=(2\pi)^3 n_0 \delta ({\bf k})+\dsf{n_0
R(\bf k)}{(\hbar^2 {\bf k}^2/2m + 2ng)^2}\,, \lab{lspeck55} \ee
in first order of $R(\bf k)$, from which the energy of the
quasiparticles follows to be $\varepsilon ({\bf p})={\bf p}^2/2m +
2ng$, where ${\bf p}=\hbar {\bf k}$. Substituting $\varepsilon ({\bf
p})$ in Eq.~\re{lspeck54} and performing its integral by parts and
using in the obtained expression $n(\varepsilon ({\bf p}))$ from
Eq.~\re{lspeck55}, one gets
\be
\rho_n=\rho_0\, \int \, \dsf{d^3 k}{(2\pi)^3} \dsf{p_z^2
R(\bf k)}{{\bf p}^2 (\hbar^2 {\bf k}^2/2m + 2ng)^2}\,,
\lab{lspeck56}
\ee
where $\rho_0=mn_0$.

It is interesting that there is the relationship
$\varepsilon_{\rm B} ({\bf p})=\varepsilon^{1/2} ({\bf p}){\bf p}/(2m)^{1/2}$ between our $\varepsilon ({\bf p})$
and the Bogoliubov quasiparticle energy $\varepsilon_B ({\bf p})$.
If we use this relation, then we obtain
\be
\rho_n=\dsf{\rho_0}{4}\,
\int \, \dsf{d^3 k}{(2\pi)^3} \dsf{{\bf p}^2\, p_z^2 R(\bf k)}{m^2
\varepsilon_B^4 ({\bf p})}\,.
\lab{lspeck57}
\ee
This expression without the prefactor $1/4$ coincides with Eq.~(19)
of Ref.~\ci{Giorgini} for the normalfluid density $\rho_{n,LR}$,
obtained within the linear response approach, if we replace $V\int
d^3 k/(2\pi)^3$ by $\sum_{\bf k}$. The prefactor $1/4$ appears from
the relation between $\varepsilon ({\bf p})$ and $\varepsilon_{\rm
B} ({\bf p})$. For a $3d$ isotropic BEC system we have
$p_x^2=p_y^2=p_z^2$ and ${\bf p}^2=3p_z^2$. Multiplying the
right-hand side of Eq.~\re{lspeck56} with 3 and canceling $3p_z^2$
and ${\bf p}^2$ in the numerator and the denominator, we obtain
Eq.~\re{lspeck42}, therefore,
$n-n_0=3\rho_{n,LR}/(4m)$~\ci{Giorgini}.

It is worth to discuss the validity to use the Landau approach for
BEC with the disorder. According to  a remark in the text
book~\ci{Pitaevskii} the Landau approach should not be applicable
for such a system. Indeed, the applied Landau derivation of the
normalfluid density presumes the validity of the quasiparticle
concept (see, for instance, Refs.~\ci{Landau1a,Landau1b}), in which
there are no collisions not only between quasiparticles but also of
last ones with the external disorder potential. More exactly,
according to this concept quasiparticles should be well defined and
their gas should be ideal.

In our case, effective quasiparticles with the mean-field energy
$\varepsilon ({\bf p})$ and the quantum state distribution at
temperature $T=0$, represented by the total density $n(\varepsilon
({\bf p}))$, appear in the system after the disorder ensemble
average. However, after this averaging the real space is homogeneous
and there is no reason for the gas of effective quasiparticles to be
not ideal. Hence, if for the conventional quasiparticles the source
of their appearance is the low temperature, here it is the
scattering of the condensate particles with the disorder and then
their excitation and departure from the condensate. This physical
conclusion naturally arises from the Landau derivation of the
normalfluid density.

\section{Summary and conclusion}
\label{sec7}

At first, we have summarized the derivation of the
autocorrelation function of the laser speckle in $1d$ and $2d$
following the seminal work of Goodman. We showed that a Gaussian
approximation of this function, proposed in some recent papers, is
inconsistent with the background of laser speckle theory. Then
we have proposed a possible experimental realization for an
isotropic $3d$ laser speckle potential and derived its corresponding
autocorrelation function. Using a Fourier transform of that
function, we calculated both condensate depletion and sound velocity
of a BEC in a weak speckle disorder within a perturbative solution
of the Gross-Pitaevskii equation. At the end, we reproduced the
expression of the normalfluid density obtained earlier within the
treatment of Landau. This physically transparent derivation showed
that condensate particles, which are scattered by disorder, form a
gas of quasiparticles which is responsible for the normalfluid
component. We have justified the validity of the Landau approach to
our BEC system with disorder.

\section{Acknowledgements}
\label{sec8}

One of the authors, B. A., thanks the Volkswagen Foundation for
partial support of the work. B. A. is also grateful to Center for International Cooperation
at the Freie Universit\"at
Berlin for its hospitality. Both authors appreciate Hagen
Kleinert and the members of his group for many discussions.


\begin{thebibliography}{0}

\bibitem{Fisher}
M. P. A. Fisher, P. B. Weichman, G. Grinstein, and D. S. Fisher,
Phys. Rev. B {\bf 40}, 546 (1989).

\bibitem{Chan}
M. H. W. Chan, K. I. Blum, S. Q. Murphy, G. K. S. Wong, and J. D. Reppy,
Phys. Rev. Lett. {\bf 61}, 1950 (1988).

\bibitem{Anderson}
P. W. Anderson,
Phys. Rev. {\bf 109}, 1492 (1958).

\bibitem{LSP}
L. Sanchez-Palencia and M. Lewenstein,
Nature Phys. {\bf 6}, 87 (2010).

\bibitem{Billy}
J. Billy, V. Josse, Z. Zuo, A. Bernard, B. Hambrecht, P. Lugan, D. Clement, L. Sanchez-Palencia, P. Bouyer, and A. Aspect,
Nature {\bf 453}, 891 (2008).

\bibitem{Roati}
G. Roati, C. D'Errico, L. Fallani, M. Fattori, C. Fort, M. Zaccanti, G. Modugno, M. Modugno, and M. Inguscio,
Nature  {\bf 453}, 895 (2008).

\bibitem{Goodman1}
J. W. Goodman,
{\it Statistical Properties of Laser Speckle Patterns}
in J. C. Dainty (Editor),
{\it Laser Speckle and Related Phenomena}
(Springer-Verlag, Berlin, 1975).

\bibitem{Boers}
D. J. Boers, B. Goedeke, D. Hinrichs, and M. Holthaus, Phys. Rev. A
{\bf 75}, 063404 (2007).

\bibitem{Kondov}
S. S. Kondov, W. R. McGehee, J. J. Zirbel, and B. DeMarco,
Science {\bf 334}, 66 (2011).

\bibitem{Pezze}
L. Pezze, M. Robert-de-Saint-Vincent, T. Bourdel, J.-P. Brantut, B. Allard, T. Plisson, A. Aspect, P. Bouyer, and L. Sanchez-Palencia,
New J. Phys. {\bf 13}, 095015 (2011).

\bibitem{Goodman2}
J. W. Goodman,
{\it Speckle Phenomena in Optics: Theory and Applications}
(Roberts and Co, Englewood, 2007).

\bibitem{Dainty}
J. C. Dainty, {\it An introductiuon to 'Gaussian' speckle} in SPIE,
Vol. 243 {\it Applications of Speckle Phenomena} (1980).

\bibitem{Pilati1}
S. Pilati, S. Giorgini, and N. Prokof'ev,
Phys. Rev. Lett. {\bf 102}, 150402 (2009).

\bibitem{Pilati2}
S. Pilati, S. Giorgini, M. Modugno, and N. Prokof'ev,
New J. Phys. {\bf 12}, 073003 (2010).

\bibitem{Piraud}
M. Piraud, L. Pezze, and L. Sanchez-Palencia,
Europhys. Lett. {\bf 99}, 50003 (2012).

\bibitem{Clement1a}
D. Clement, A. F. Varon, M. Hugbart, J. A. Retter, P. Bouyer, L. Sanchez-Palencia, D. M. Gangardt, G. V. Shlyapnikov, and A. Aspect,
Phys. Rev. Lett. {\bf 95}, 170409 (2005).

\bibitem{Clement1b}
D. Clement, P. Bouyer, A. Aspect, and L. Sanchez-Palencia,
Phys. Rev. A {\bf 77}, 033631 (2008).

\bibitem{Chen1}
Y. P. Chen, J. Hitchcock, D. Dries, M. Junker, C. Welford, and R. G. Hulet,
Phys. Rev. A {\bf 77}, 033632 (2008).

\bibitem{Chen2}
D. Dries, S. E. Pollack, J. M. Hitchcock, and R. G. Hulet,
Phys. Rev. A {\bf 82}, 033603 (2010).

\bibitem{Chen3}
C. Fort, L. Fallani, V. Guarrera, J. E. Lye, M. Modugno, D. S. Wiersma, and M. Inguscio,
Phys. Rev. Lett. {\bf 95}, 170410 (2005).

\bibitem{Chen4}
M. Robert-de-Saint-Vincent, J. -P. Brantut, B. Allard, T. Plisson, L. Pezze, L. Sanchez-Palencia, A. Aspect, T. Bourdel, and P. Bouyer,
Phys. Rev. Lett. {\bf 104}, 220602 (2010).

\bibitem{Clement2}
D. Clement, A. F. Varon, J. A. Retter, L. Sanchez-Palencia, A. Aspect, and P. Bouyer,
New J. Phys. {\bf 8}, 165 (2006).

\bibitem{Leushacke}
L. Leushacke and M. Kirchner, J. Opt. Soc. Am. A {\bf 7}, 827
(1990).

\bibitem{Middleton}
D. Middleton, {\it Introduction to Statistical Communication Theory}
(McGraw Hill, New York, 1960).

\bibitem{Kuhn}
R. C. Kuhn, O. Sigwarth, C. Miniatura, D. Delande, and C. A. Muller,
New J. Phys. {\bf 9}, 161 (2007).

\bibitem{Modugno}
M. Modugno,
Phys. Rev. A {\bf 73}, 013606 (2006).

\bibitem{Huntley}
J. M. Huntley,
Appl. Opt. {\bf 28}, 4316 (1989).

\bibitem{Makse1}
H. A. Makse, S. Havlin, M. Schwartz, and H. E. Stanley,
Phys. Rev. E {\bf 53}, 5445 (1996).

\bibitem{Makse2}
P. R. Kramer, O. Kurbanmuradov, and K. Sabelfeld,
J. Comput. Phys. {\bf 226}, 897 (2007).

\bibitem{Sucu}
S. Sucu,  S. Aktas, S. E. Okan, Z. Akdeniz, and P. Vignolo,
Phys. Rev. A {\bf 84}, 065602 (2011)

\bibitem{Jendrzejewski}
F. Jendrzejewski, A. Bernard, K. Mueller, P. Cheinet, V. Josse, M.
Piraud, L. Pezze, L. Sanchez-Palencia, A. Aspect, P. Bouyer, Nature
Phys. {\bf 8}, 398 (2012).

\bibitem{Krumnow}
C. Krumnow and A. Pelster,
Phys. Rev. A {\bf 84}, 021608(R) (2011).

\bibitem{Huang}
K. Huang and H.-F. Meng,
Phys. Rev. Lett. {\bf 69}, 644 (1992).

\bibitem{Falco}
G.M. Falco, A. Pelster, and R. Graham,
Phys. Rev. A {\bf 75}, 063619 (2007).

\bibitem{Giorgini}
S. Giorgini, L. Pitaevskii, and S. Stringari,
Phys. Rev. B {\bf 49}, 12938 (1994).

\bibitem{Landau1a}
L. Landau, J. Phys. USSR, {\bf 5}, 71 (1941).

\bibitem{Landau1b}
E. M. Lifshitz and L. P. Pitaevskii, {\it Statistical Physics.
Theory of the Condensed State} (Elsevier Ltd., Amsterdam 1980).

\bibitem{Landau2}
L. D. Landau and E. M. Lifshitz, {\it Quantum Mechanics.
Non-Relativistic Theory} (Elsevier Ltd., Amsterdam 1977).

\bibitem{Pitaevskii}
L. Pitaevskii and S. Stringari, {\it Bose-Einstein Condensation}
(Clarendon Press, Oxford, 2003), p.66.
\end{thebibliography}
\end{document}